# Cryogenic Loss Monitors with FPGA TDC Signal Processing

A.Warner[a*], J. Wu[a]

*Fermi National Accelerator Laboratory, P.O.Box 500, Batavia IL 60510, U.S.A.*

**Abstract**

Radiation hard helium gas ionization chambers capable of operating in vacuum at temperatures ranging from 5K to 350K have been designed, fabricated and tested and will be used inside the cryostats at Fermilab's Superconducting Radiofrequency beam test facility. The chamber vessels are made of stainless steel and all materials used including seals are known to be radiation hard and suitable for operation at 5K. The chambers are designed to measure radiation up to 30 kRad/hr with sensitivity of approximately 1.9 pA/(Rad/hr). The signal current is measured with a recycling integrator current-to-frequency converter to achieve a required measurement capability for low current and a wide dynamic range. A novel scheme of using an FPGA-based time-to-digital converter (TDC) to measure time intervals between pulses output from the recycling integrator is employed to ensure a fast beam loss response along with a current measurement resolution better than 10-bit. This paper will describe the results obtained and highlight the processing techniques used.

## 1. Introduction

Loss monitors are typically one of the main diagnostic devices used in high beam power super-conducting RF accelerators for protecting the machines from beam induced damage and radiation damage; as high gradient RF structures can produce field emission that results in so called "dark current". The dark current related radiation that results can cause significant damage to accelerator components. Although the placements of loss monitors are critical, most facilities do not cover cold sections of the machine with loss monitors. To address these issues a Cryogenic Loss Monitor (CLM) ionization chamber capable of operation in the cold sections of a cryomodule has been developed and will be installed and

―――――――

* Corresponding author. Tel.:1-630-840-6119; fax:1-630-840-4721.
  *E-mail address*: warner@fnal.gov.

tested at the Superconducting RF accelerator test facility currently under construction at Fermilab [1]. The monitor electronics have been optimized to be sensitive to DC losses and the signals from these devices will be used to study and quantify dark current losses in particular. In order to increase the resolution bandwidth and the response time of the devices a new scheme which uses a Field Programmable Gate Array (FPGA) based Time-to-Digital converter method was implemented [2] instead of a standard counter or digitizer method. This potentially renders these monitors as useful devices for both dark current monitoring and machine protection.

## 2. Cryogenic Loss Monitor Design

The cryogenic loss monitors are all metal ionization chambers with the exception of the ceramic insulator for the signal electrode, see Fig 1. The devices filled with 120 cm$^3$ of Helium gas at 1.0 bar pressure. The all metal design makes them intrinsically radiation hard and suitable for operation from 5 Kelvin to 350 Kelvin. The calibration "$S$" of the monitors is almost completely determined by the volume "$V$" and type of gas:

$$S \approx V.\rho.e/E_{ion}$$

Here $\rho$ is the gas density, $e$ is the electron charge and $E_{ion}$ is mean energy deposition to create electron-ion pairs. The housing of the chambers is held at a negative potential of -95 V and is kept well below the minimum breakdown voltage of 156 Volts for Helium. Negative charges are collected on the centre electrode of the device. The device essentially functions as a dose rate monitor. The signal current which is proportional to the dose rate is initially processed by a recycling integrator which acts as a current to frequency converter with a wide dynamic range. The recycling integrator consist of a charge integrating amplifier with a 0.5 pF (nominal) capacitance followed by a discriminator, which senses when the capacitor is fully charged to produce a fixed width (1.2 µs) discharge pulse. The discharge pulses are mirrored unto a NIM output (-0.80V into 50 Ω) and can be transferred via long cables. The charge per pulse is 1.63 pC or 238 µR at 1 atmosphere.

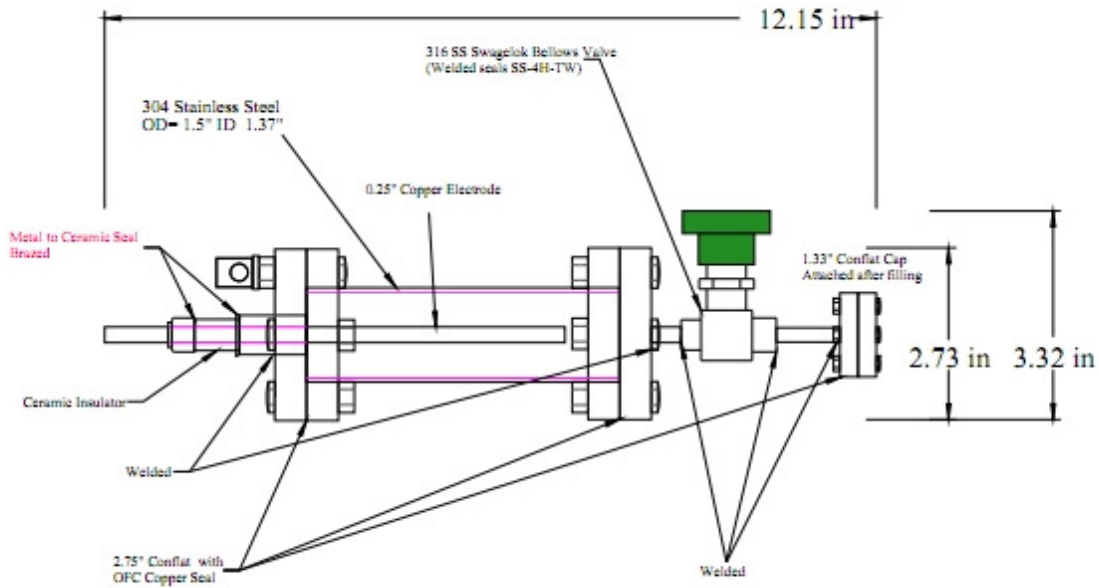

## 3. FPGA-TDC Signal Processing

In typical digitization/readout schemes as shown in Fig. 2(a), a counter is utilized to accumulate the number of pulses generated by the recycling integrator to adequately digitize the total charge. In order to calculate current with reasonable resolution (e.g., 7-8 bits), hundreds of pulses must be accumulated which corresponds to either a long sampling period, or a very low sampling rate. For example, to achieve 7-bit resolution, the sampling period corresponds to 128 pulses when input current is at its upper limit. Such schemes provide a total dosage of the radiation over long periods but are not fast enough for accelerator beam protection.

In our new scheme shown in Fig. 2(b), the recycling integrator output is sent to an FPGA in which a time-to-digital converter (TDC) is implemented. The TDC is utilized to measure the time intervals between the pulses output from the recycling integrator. When the input current is not too high, the recycling integrator pre-charges the capacitor with a fix width pulse in each cycle so that a known constant charge "$Q$" is stored in the capacitor. When the time difference "$dt$" between two leading edges of the output pulses is measured, the average current "$I$" during the time interval is approximately $<I> = Q/dt$. The TDC is based on a multi-sampling scheme in which the input transition is sampled with four different phases of the system clock. With system clock rates of 250 MHz and four-phase sampling, a 1-ns time measurement resolution can be achieved. Using this method, a sample measurement of the current can be made with good timing resolution (>10 bits) between each pulse. This effectively increases the sampling rates by hundreds of times for the same recycling integrator front-end electronics which provides a fast response to the beam loss and is potentially suitable for accelerator protection applications. More-over, the method is also self-zero-suppressed, i.e., it produces more data when the beam loss is high while it produces significantly less data when the beam loss is low.

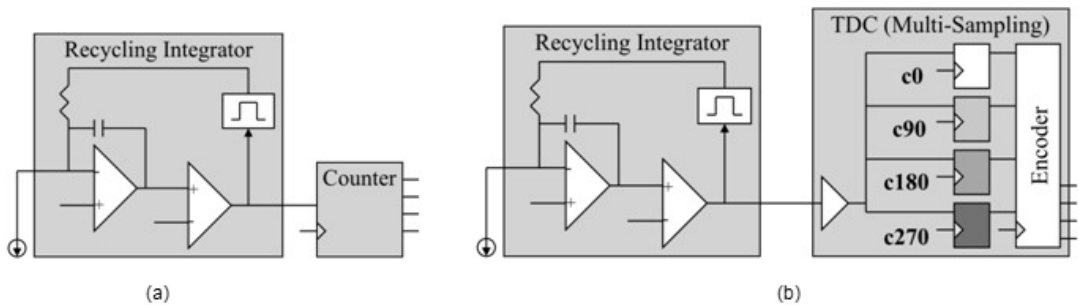

Fig. 2 (a) typical digitization scheme with counter; (b) Novel digitization scheme using FPGA-TDC

## 4. FPGA-TDC Test Bench Measurements

Bench top test results are shown in Fig. 3. In Fig. 3(a), the upper oscilloscope trace represents the input current on a scale of 100 nA/div and the lower trace is the output of the recycling integrator with a calibration of approximately 2 pC/pulse. If the pulses are accumulated in a counter, about 60 pulses are counted for the entire time frame indicating a total charge of 120 pC and the measurement resolution is approximately 6 bits. With the TDC, differences of time between pulses are measured that give a current measurement sample per pulse as shown in Fig. 3(b). It can be seen that when the input current becomes large, it is detected almost immediately as the time interval between two pulses becomes short.

With the TDC as the digitization device, challenges to the analog circuit design in the recycling integrator can be reduced. For example, when the input current is large, the pre-charge pulse in the integrator may become wider which causes more total charge to be stored in the capacitor. The TDC can measure the transition times of both edges of the output pulse and the width of the pulse can be considered to be proportional to the total charge stored in each cycle. With tolerance of the pulse width, the dynamic range of the recycling integrator can be further extended.

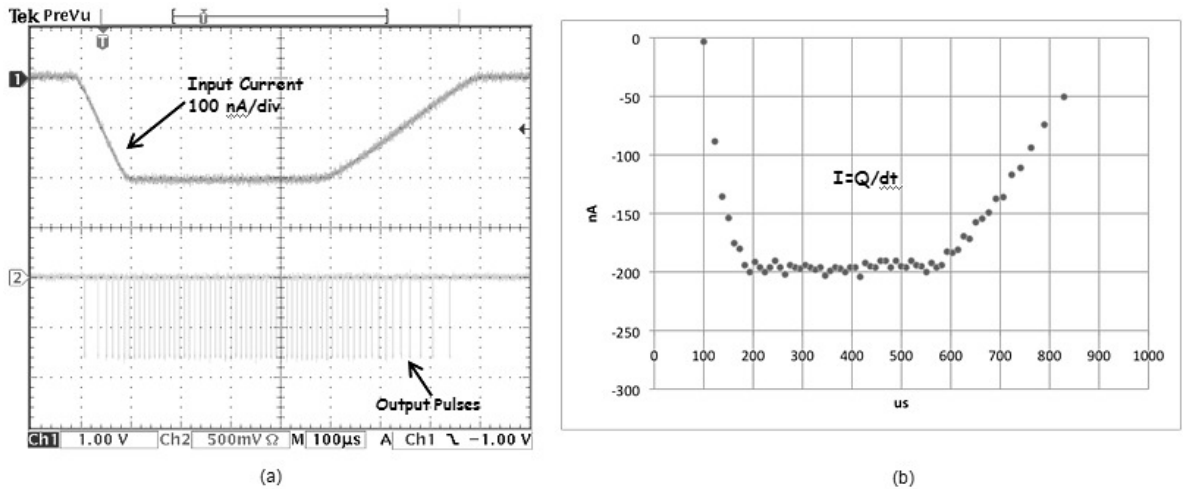

Fig. 3 Bench top test results: (a) Oscilloscope traces of the input current and the output pulses of the recycling integrator; (b) Measured current from the time difference between the pulses

It is interesting to compare features of this combined recycling integrator and TDC scheme with the regular ADC scheme. In the regular ADC circuit, an amplifier circuit with sufficient gain and careful grounding is necessary to interface with the low current and noise sensitive signal source of the ionization chamber; whereas the recycling integrator is designed to interface directly with such low current sources. Regular ADCs provide data samples at a constant rate and fix resolution. This scheme, on the other hand, provides rapid response with relatively low measurement resolution when the input current is high, while it provides slower measurements for low current with higher resolution when the time interval between pulses is longer. This is a good fit for our accelerator beam loss monitor since both rapid high beam loss protection and precise measurement of the dark current are needed.

### 5. Cryogenic Test set-up

Measure where made at the Femilab Horizontal Test Stand (HTS) [3] during cold cavity conditioning (no beam) and at the A0 Photo-injector [4] test facility (room temperature). The schematic layout of the loss monitor setup for cryogenic testing is shown in fig. 4(a) below. The cross section illustrates the divide between the cold vacuum side of the vessel and air (outside). The ionization current is carried via its own cable to the recycling integrator electronics box which also provides both the bias voltage and the recycling integrator's pulsed output signal. This output pulse which is first mirrored to a NIM output is then converted to a Low Voltage Differential Signal (LVDS) before processing by the FPGA-TDC. A simplified schematic view of the recycling integrator electronics is also shown below in Fig 4(b).

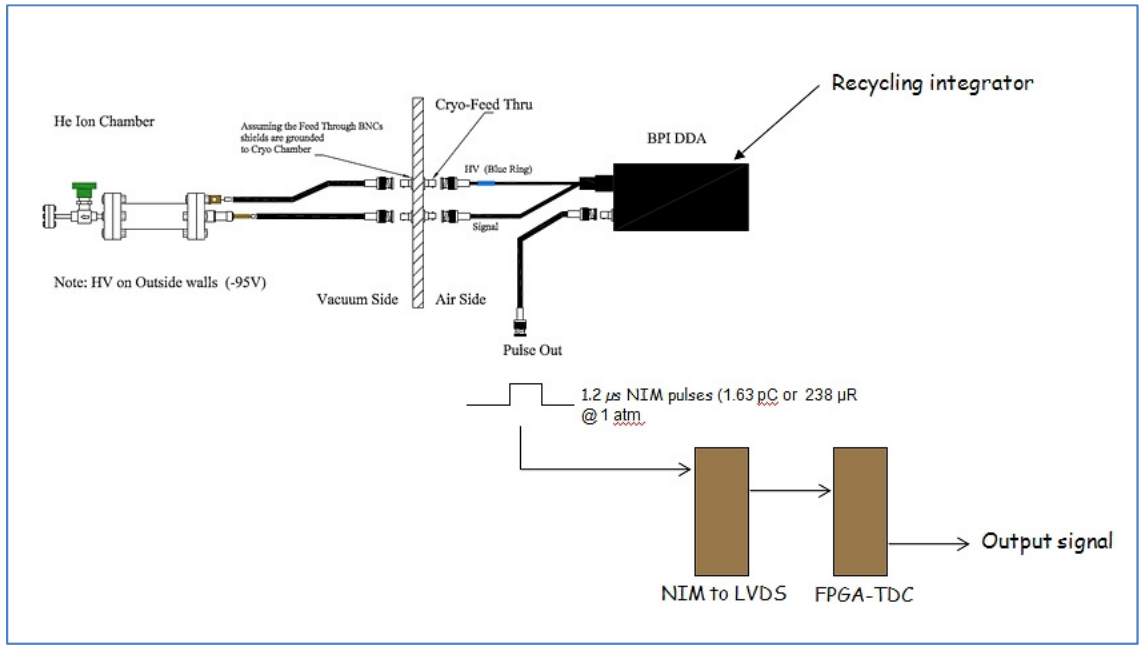

Fig. 4 (a) Schematic diagram Cryogenic Loss Monitor connection and signal path

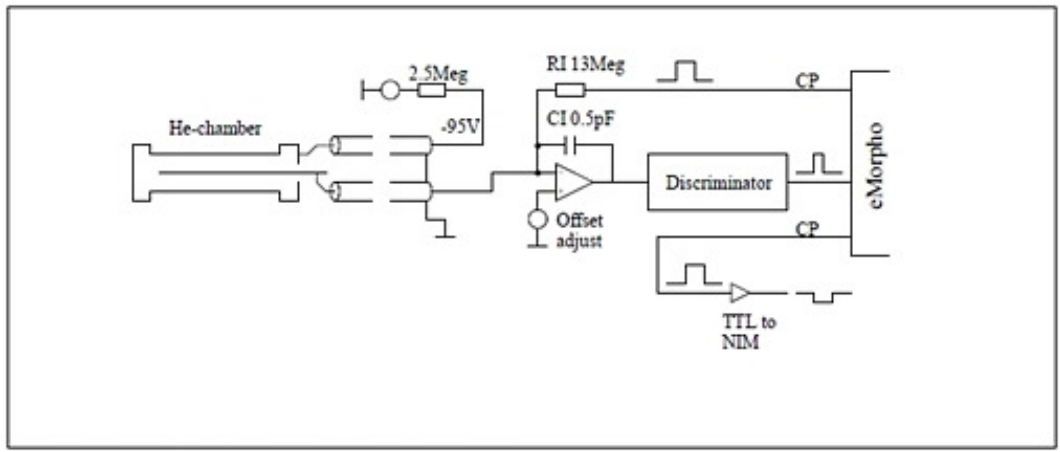

Fig. 4 (b) Schematic view of recycling integrator electronics

Fig. 5(a) shows the location of the cryogenic loss monitor relative to a 9 cell superconducting RF cavity under test at HTS and Fig. 5(b) shows the proposed installation in a cryomodule for future test.

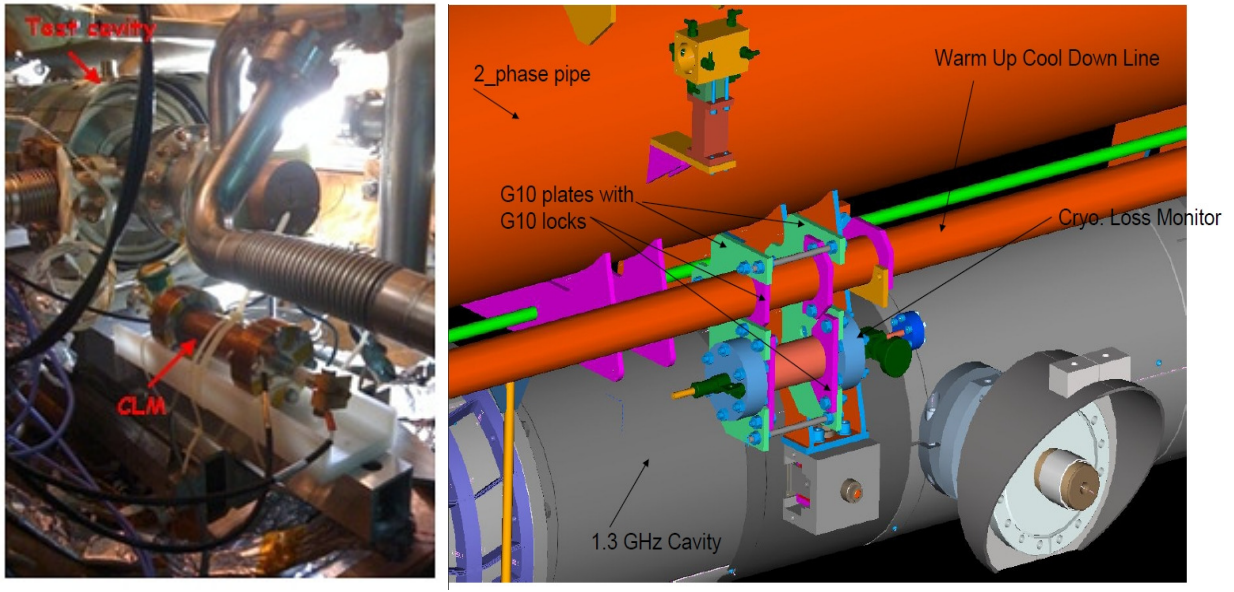

Fig. 5 (a) CLM positioned in the Horizontal Test Stand; (b) Propose installation in a cryomodule

**6. Dark Current Measurement result.**

Background dark current losses were measured at the A0-Photo-injector test accelerator Fig. 6. The lower image shows the scope trace from a 40 µs pulse of the RF with no photo-electrons present in the machine. The upper trace is the output from the CLM recycling integrator electronics and corresponds to about 400 nA (~200 mR).  Similar measurements at the HTS under cold conditions produce dark current levels of the order 150 mR and were consistent with other loss monitors in the system.

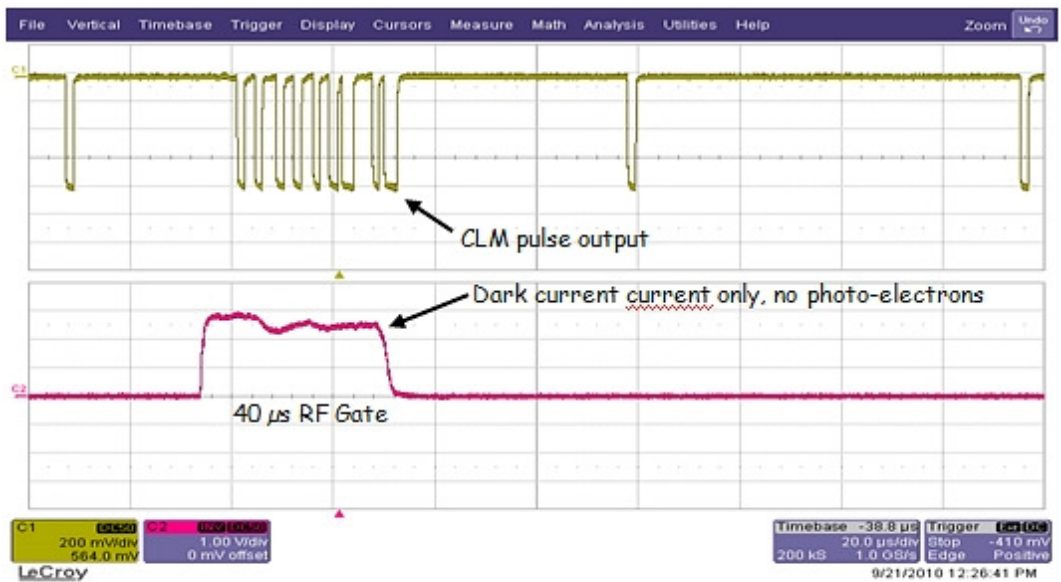

Fig. 6

## 7. Summary

Unlike a regular ADC device that produces the same amount of data no matter what the input current, this scheme is self zero-suppressed, i.e., it produces more data when the beam loss (and therefore the input current) is high and produces significantly less data when there is no beam loss. This feature further simplifies design of the readout system and the data analysis software.

## Acknowledgements

The authors would like to thank Bridgeport Instruments LLC, for they customized work on the fabrication and design of the loss monitors and recycling integrator electronics; also for working with us with the customizations that were necessary for us to implement the FPGA-TDC electronics.